\documentclass[aps,pra,twocolumn,showpacs]{revtex4-1}
\DeclareMathAlphabet{\mathpzc}{OT1}{pzc}{m}{it} 
\usepackage{mathrsfs}
\usepackage{tabularx}
\usepackage{dcolumn}
\newcolumntype{C}{>{\centering\arraybackslash}X}
\newcolumntype{R}{>{\raggedleft\arraybackslash}X}
\newcolumntype{L}{>{\raggedright\arraybackslash}X}
\usepackage{graphics}
\usepackage{dcolumn}
\usepackage{bm}
\usepackage[ps2pdf]{graphicx}
\usepackage[dvips]{rotating}
\usepackage[latin1]{inputenc}
\usepackage{dcolumn}
\usepackage{tabularx}
\usepackage{amsmath}
\usepackage{amssymb}
\begin{document}

\title{Positive-energy spectra of atomic hydrogen in a magnetic field with an adiabatic-basis-expansion method}
\author{L. B. Zhao$^1$}
\author{K. D. Wang$^2$}
\author{K. Bartschat$^3$}
\affiliation{$^1$Department of Physics and Astronomy, Guizhou University, Guiyang 550025, China}
\affiliation{$^2$School of Physics, Henan Normal University, Xinxiang, 453007, China} 
\affiliation{$^3$Department of Physics and Astronomy, Drake University, Des Moines, Iowa, 50311, USA}
\date{\today}
\begin{abstract}
The problem of photo\-ionization of atomic hydrogen in a white-dwarf-strength magnetic field is 
revisited to understand the existing discrepancies in the positive-energy spectra obtained by  
a variety of theoretical approaches reported in the literature. Oscillator strengths 
for photo\-ionization are calculated with the adiabatic-basis-expansion method developed by 
Mota-Furtado and O'Mahony [Phys.\ Rev.\ A~{\bf 76}, 053405 (2007)]. A comparative study is 
performed between the adiabatic-basis-expansion method and our previously developed 
coupled-channel theory [Phys.\ Rev.\ A~{\bf 94}, 033422 (2016)]. A detailed analysis of the positive-energy 
spectra obtained here and those from other theoretical approaches   
shows that the adiabatic-basis-expansion method can produce more accurate positive-energy 
spectra than other reported approaches for low field strengths. 
\end{abstract}
\maketitle 
\newpage
\section{Introduction}\label{sec:Intro} 
Understanding the behavior of atoms in the presence of magnetic fields has been a 
subject of considerable interest, since Zeeman experimentally discovered the splitting 
of atomic spectral lines in a magnetic field at the end of the 19th century~\cite{Garstang}. 
From then on, a great deal of effort has been devoted to the study of magnetized atomic 
systems. Many early attempts in this aspect were made within the framework of classical 
theory, before the establishment of quantum mechanics gave researchers a 
deeper insight into the dynamics of magnetized atoms. Perturbation theory achieved 
prominent success in quantitatively explaining the splitting of spectral lines of hydrogen 
atoms in a weak magnetic field in the early stages of quantum mechanics. 
However, it took a long time to understand the behavior of atoms in a strong 
magnetic field, i.e., the so-called quadratic Zeeman effect. 

A spectroscopic experiment on highly excited barium atoms in a strong magnetic field 
of laboratory strength, reported by Garton and Tomkins~\cite{Garton} in 1969, 
stimulated interest in the quadratic Zeeman effect. Such interest was reinforced by a 
growing number of subsequent experiments on other strongly-magnetized atoms 
(see, e.g., Refs.~\cite{Holle,Iu} and references therein). The regular quasi-Landau 
resonance structures observed in the negative-energy region close to the ionization 
thresholds were interpreted by the semiclassical theory developed by Du and 
Delos~\cite{Du}. The diamagnetic Rydberg spectra of the alkali-metal atoms in the 
positive-energy region were also reproduced by fully quantum-mechanical approaches 
\cite{OMahony,Watanabe}. Undoubtedly, the study on the structures and dynamics of 
atoms in the presence of magnetic fields has made significant headway up to the 1990s. 

The discovery of superstrong magnetic fields in white-dwarf stars with field strengths 
$10^2-10^5$ T and neutron stars with field strengths $10^7-10^9$ T 
reinforced interest in theoretical investigations of atoms in the presence of high magnetic 
fields~\cite{Garstang,Ferrario2015}. In astronomy and 
astrophysics, spectra from highly accurate theories are indispensable for determining 
the size of magnetic fields in the atmospheres of white-dwarf and neutron stars. So far, 
many efforts have been dedicated to developing non\-perturbative theories and 
numerical methods to calculate properties of magnetized atoms (see, e.g., Ref.~\cite{Ruder1994} 
and references therein). A multi\-configuration Hartree-Fock method was developed to 
calculate the properties of magnetized hydrogen atoms~\cite{Rosner1984,Forster1984}. 
This method provides a tool to model spectral lines of bound-bound transitions for 
hydrogen atoms in magnetic fields of both white dwarfs and neutron stars. More recently, 
a finite-basis-size method~\cite{Zhao2007} was extended to calculate discrete 
spectral lines of the Lyman, Balmer, and Paschen series for magnetized hydrogen atoms 
\cite{Zhao2019,Zhao2020}. 

At this time, it has become possible to model discrete spectra of hydrogen 
atoms in an arbitrary magnetic field, based on the above-mentioned theoretical approaches. 
While this represents significant progress in astrophysical applications, 
theories for describing multi\-electron atoms in a magnetic field are still very 
scarce, and far from satisfying the demands of analyzing the spectra observed in magnetic 
white dwarfs and neutron stars. The origin of the problem can be attributed 
to the difficulties of treating electron correlations in a strong magnetic field. Very recently, 
the Zeeman splitting lines of multi\-electron atoms, such as He, Ca, Mg, Fe, and Na, 
were discovered in the atmospheres of magnetic white dwarfs~\cite{Ferrario2020}. 
Consequently, it is inevitable to develop theories and numerical methods to calculate 
the properties of strongly magnetized multi\-electron atoms. 

As mentioned above, identifications of discrete spectra of some magnetized atoms have been
successful. However, the analysis of continuum spectra observed from magnetic celestial 
objects suffers from the lack of accurate photo\-ionization cross sections. To our knowledge, no theoretical 
study on photo\-ionization of highly magnetized multi\-electron atoms has yet been reported in the 
literature, and only a few theoretical approaches have been presented for describing bound-free 
transitions of hydrogen atoms in magnetic fields with white-dwarf-field strengths. Alijah 
{\it et al.}~\cite{Alijah} developed a theoretical approach based on multi\-channel 
quantum defect theory to describe bound-free transitions for magnetized hydrogen 
atoms. The wave functions for the continuum states were obtained by numerically solving the 
coupled Schr\"odinger equations. The photo\-ionization spectrum from the ground state 
was published for a magnetic field of 2,000 T, and a Rydberg series of resonance states 
was identified. The authors expressed confidence that an extension to field strengths lower than 
2,000 T was possible with some modifications of their approach. 

Delande {\it et al.}~\cite{Delande} presented a complex-rotation method to calculate 
positive-energy spectra of hydrogen atoms in a magnetic field. The photo\-ionization 
spectrum from the ground state in a magnetic field of 23,500~T was reported. This 
spectrum has become the benchmark for testing theories of 
bound-free transitions in a strong magnetic field. Merani {\it et al.}~\cite{Merani} 
developed a complex-coordinate method similar to that of Delande {\it et al.} to 
produce the vast amount of data on photo\-ionization cross sections as a function of 
field strengths and energies that are required for the analysis of spectra observed in magnetic 
celestial objects. Balmer and Paschen bound-free opacities were determined from  
the obtained cross sections. Zhao and Stancil~\cite{Zhao2006} presented a computational 
scheme based on the complex-rotation method using a mixed Slater-Landau basis to yield 
continuum spectra of magnetized hydrogen atoms. In order to describe atomic hydrogen 
diamagnetism, Wang and Greene~\cite{Wang} combined $R$-matrix theory with 
multi\-channel quantum defect theory to construct a theoretical approach to calculate 
photo\-ionization cross sections of hydrogen atoms in magnetic fields with strengths 
of $10^3-10^4$ T. However, this approach did not reproduce 
the continuum spectrum at 2,000~T published by Alijah {\it et al.}~\cite{Alijah}. 

The discrepancies of the continuum spectrum at 2,000~T from the two theoretical studies 
mentioned above motivated us to further explore the dynamics of bound-free transitions 
in a strong magnetic field. A coupled-channel theory was developed for this 
purpose~\cite{Zhao2016}, but this theory could reproduce neither the spectrum of 
Alijah {\it et al.}~\cite{Alijah} nor that of Wang and Greene~\cite{Wang}. 
This unexpected result stimulated us to revisit the problem of photo\-ionization 
of atomic hydrogen in a white-dwarf-strength magnetic field. In the present work, we 
adopt the adiabatic-basis-expansion method established by Mota-Furtado and O'Mahony 
\cite{OMahony2007} to perform such an investigation. Our purpose  
is to elucidate potential reasons for the discrepancies in the continuum spectra at 2,000~T 
obtained in the various calculations. 

This manuscript is organized as follows. Section~\ref{sec:Theory} is devoted to sketching the 
adiabatic-basis-expansion method developed by Mota-Furtado and O'Mahony 
\cite{OMahony2007}, which is used to study the photo\-ionization of hydrogen atoms in a 
strong magnetic field in the current paper. In Sect.~\ref{sec:Results}, the adiabatic-basis-expansion 
method is applied to calculating continuum spectra of magnetized hydrogen atoms. 
A comparative study of these spectra close to the ionization thresholds is performed 
between this method and our previously developed coupled-channel theory  
\cite{Zhao2016}. The predicted continuum spectra are also compared to those from 
other theoretical approaches. A detailed analysis of the existing discrepancies 
among the continuum spectra obtained by the different theoretical methods is presented 
in this section. Section~\ref{sec:Summary} summarizes the results of the current study and our main 
conclusions regarding photo\-ionization of hydrogen atoms in strong magnetic fields. 

Atomic units are used throughout this paper unless otherwise noted. 

\section{Sketch of the theoretical method}\label{sec:Theory}
The adiabatic-basis-expansion method developed to study photo\-ionization of hydrogen 
atoms in the presence of magnetic fields was formulated by Mota-Furtado and 
O'Mahony in Ref.~\cite{OMahony2007}, where more theoretical details  
can be found. Here we only outline the general flow of arguments and point out significant features 
relevant to the current work. 

Suppose a hydrogen atom in some initial state 
is placed in a magnetic field $B$ pointing along the $z$ axis. The Hamiltonian $H$ for this 
atomic system is written in the form 
\begin{equation} 
\widehat{H}=-\frac 12\nabla ^2-\frac 1r + 
\frac{\gamma}2({\hat \ell}_z + 2{\hat s}_z) + \frac18{\gamma^2}r^2\sin^2\theta,
\label{full_Hamiltonian}
\end{equation}
where $\gamma=B/B_0$ is the magnetic field strength in atomic units, $i.e.$, in 
multiples of $B_0 \approx 2.35 \times 10^5$ T, ${\hat \ell}_z$ and ${\hat s}_z$ are 
the $z$ components of the orbital and spin angular momenta, respectively, the third 
term (linear in $\gamma$) is the paramagnetic potential, and the fourth term (quadratic 
in $\gamma$) is the diamagnetic potential.  For this atomic system, the orbital angular 
momentum is not a good quantum number specified by $\ell$, but its projection on 
the $z$ axis is a good quantum number specified by $m$. Furthermore, the $z$ parity 
of the eigenstates, denoted by $\pi_z$ below, is conserved. Here $\pi_z$ and $m$ are adopted to 
identify the hydrogen atomic states, labeled by $m^{\pi_z}$, in a magnetic field.  

When such an atom is irradiated by a beam of polarized light, it may absorb a photon 
and then be ionized. The electron wave produced by the ionization process propagates 
from the inner to the outer region. The entire configuration space is divided into multiple 
radial sectors with radii $a \to a_1 \to a_2, \cdot \cdot \cdot, \to a_{N-1} \to a_N=b$, 
where $N$ is the number of the sectors, while $a$ and $b$ are the inner and 
outer radius, respectively. To begin with, the local adiabatic basis is constructed in each sector, and 
then the radial wave functions or the $R$-matrix are propagated from the inner region 
to the asymptotic region, sector by sector. Finally, a two-dimensional matching to 
the $R$-matrix in the asymptotic region, where the solutions of the coupled Schr\"odinger 
equations in cylindrical coordinates are attainable, is performed to extract the reactance
matrices. Throughout this paper, we will use as much as possible the same symbols as in Ref.~\cite{OMahony2007}, 
where the details of the adiabatic-basis-expansion method are given. 

\subsection{Adiabatic eigenstates}
The first step to develop the adiabatic-basis-expansion method is to construct the local 
adiabatic basis set. The adiabatic Hamiltonian $\widehat{H}_{ad}$ is written as   
\begin{equation}
\widehat{H}_{ad}(r_a^n; \theta, \phi) = \frac{\hat{\ell}^2}{2(r_a^n)^2} -
\frac{1}{r_a^n} + \frac18\gamma^2(r_a^n)^2\sin^2\theta,
\label{adiabatic_Hamiltonian}
\end{equation}  
where $\hat{\ell}$ is the orbital angular momentum operator and $r_a^n$ is the radius 
of the $n$th sector lying in the interval $a_{n-1} < r_a^n < a_n$. It is often 
taken as the midpoint of the sector. By selecting a basis set of spherical harmonics 
and calculating each matrix element of the adiabatic Hamiltonian in the basis set 
selected, one can diagonalize the resulting adiabatic Hamiltonian matrix and thereby 
obtain the adiabatic eigenstates $\varphi_{\lambda}(r_a^n; \theta, \phi)$ and 
eigenvalues $U_{\lambda}(r_a^n)$. The adiabatic eigenvalue curves obtained can 
be plotted as a function of the radius~$r$. Such curves illustrate equal energy 
spacing of Landau states at large~$r$, which is regarded as the asymptotic region
where the two-dimensional matching can begin. 

\subsection{$\bm{R}$-matrix propagation}
A basis set of orthogonal radial functions is essential in order to calculate the radial 
wave functions of the Hamiltonian (\ref{full_Hamiltonian}) in each sector. Such a 
basis set was defined by Mota-Furtado and O'Mahony~\cite{OMahony2007} in 
terms of Legendre polynomials, while the full basis set of the Hamiltonian (\ref{full_Hamiltonian}) 
consists of the product of the orthogonal radial functions, denoted by $f_j(r)$ with 
$j=1, 2, 3, \dotso$, and the adiabatic functions in one sector, 
$\varphi_{\lambda}(r_a^n; \theta, \phi)$. With the basis set obtained, the solution 
of the Hamiltonian equation 
\begin{equation}
(\widehat{H} - \varepsilon) \Psi = 0 
\label{full_Hamiltonian_no_Block_local}
\end{equation}  
is practicable by matrix diagonalization. However, 
$\widehat{H}$ is not Hermitian in the individual sectors, due to non\-vanishing surface terms. Equation 
(\ref{full_Hamiltonian_no_Block_local}), therefore, has to be revised to ensure the hermiticity 
of the Hamiltonian operator involved. 

Adding $\widehat{L}\Psi$ on both sides of the above equation, 
Eq.~(\ref{full_Hamiltonian_no_Block_local}) is rewritten as  
\begin{equation}
(\widehat{H} + \widehat{L} - \varepsilon )\Psi=\widehat{L}\Psi.
\label{full_Hamiltonian_local}
\end{equation}  
Here $\widehat{L}$ is the Bloch operator introduced in Ref.~\cite{OMahony2007} 
to guarantee that $\widehat{H}$ + $\widehat{L}$ is Hermitian.   
The above equation can be formally solved yielding  
\begin{equation}
\Psi=(\widehat{H} + \widehat{L} - \varepsilon)^{-1}\widehat{L}\Psi.
\label{full_Hamiltonian_local2}
\end{equation}  
Here $(\widehat{H} + \widehat{L} - \varepsilon)^{-1}$ is the Green's function 
given by 
\begin{equation}
(\widehat{H} + \widehat{L} - \varepsilon)^{-1} = \sum_k 
\frac{|\psi_k \rangle \langle \psi_k |} {E_k-\varepsilon},   
\label{Green_function}
\end{equation}
where $\psi_k$ and $E_k$ are eigenfunctions and eigenvalues of the operator 
$\widehat{H} + \widehat{L}$ within one sector. They are obtained  in each sector by 
diagonalizing the matrix equation of $\widehat{H} + \widehat{L}$ with matrix 
elements calculated in the basis set $\{f_j(r)\varphi_{\lambda}(r_a^n; \theta, \phi)\}$. 
Then each $\psi_k$ in the $n$th sector is expressed as
\begin{equation}
\psi_k = \sum_{jk} c_{j\lambda}^k \frac1rf_j(r)\varphi_{\lambda}(r_a^n; \theta, \phi).
\label{homogenous_wave_function}
\end{equation}
Substituting Eq.~(\ref{Green_function}) into Eq.~(\ref{full_Hamiltonian_local2}) yields
\begin{equation}
\Psi = \sum_k  \frac{|\psi_k \rangle \langle \psi_k | \widehat{L} |\Psi \rangle} 
{E_k-\varepsilon}.
\label{continuous_wave_function}
\end{equation}

The total continuum wave function $\Psi$ at any energy in the $n$th sector can 
be formally written as 
\begin{equation}
\Psi = \sum_{\lambda}\frac1rF_{\lambda}(r)\varphi_{\lambda}(r_a^n; \theta, \phi).
\label{linearly_independent_solution_in_spherical_coordinate}
\end{equation}
The right-hand sides of the above two equations are equal. 
Substituting the Bloch operator given in Ref.~\cite{OMahony2007} and Eq.~(\ref{homogenous_wave_function}) 
into the resulting equation, multiplication of 
the equation by $\varphi_{\lambda^{\prime}}^*(r_a^n; \theta, \phi)$, and 
integration over $\theta$ and $\phi$ yields equations to relate the functions and 
their derivatives at the boundaries of the $n$th sector. In 
compact matrix notation, we write these equations as 
\begin{equation}
F(a_{n-1}) = {\mathscr R}_2^nF^{\prime}(a_n) - 
{\mathscr R}_1^nF^{\prime}(a_{n-1});
\label{Fmatrix_1}
\end{equation} 
\begin{equation}
F(a_{n}) = {\mathscr R}_4^nF^{\prime}(a_n) - 
{\mathscr R}_3^nF^{\prime}(a_{n-1}).
\label{Fmatrix_2}
\end{equation} 
Here ${\mathscr R}_i^n$ with $i=1, 2, 3, 4$ are the sector $R$ matrices 
defined in Ref.~\cite{OMahony2007}, where $\bm{r}_i^n$ is used instead of 
${\mathscr R}_i^n$. While one can propagate the radial wave 
functions from one sector to its adjacent sector using the above two equations, 
it is more convenient to propagate the $R$ matrix, which relates the radial 
wave function and its derivative, $F(a_n)=R(a_n)F'(a_n)$. The relationship 
between the $R$ matrices on the inner and outer boundaries of the $n$th 
sector is derived from Eqs.~(\ref{Fmatrix_1}) and (\ref{Fmatrix_2}) as
\begin{equation}
R(a_n)={\mathscr R}_4^n-{\mathscr R}_3^n[{\mathscr R}_1^n+
R(a_{n-1})]^{-1}{\mathscr R}_2^n.
\label{R_matrix}
\end{equation} 
It should be emphasized that $R(a_n)$ and $R(a_{n-1})$ are represented in the 
same adiabatic basis set as the sector $R$ matrices. Since the adiabatic basis 
varies from one sector to another, it is essential to change the basis representation 
of the $R$ matrix for its propagation. The matrix elements of the transformation 
matrix are 
\begin{equation}
(T^{n-1,n})_{\lambda \lambda'} = \langle \varphi_{\lambda}(r_a^n; \theta, \phi) 
| \varphi_{\lambda^{\prime}}(r_a^n; \theta, \phi)  \rangle.  
\label{Transition_matrix_element}
\end{equation}
The $R$ matrix transformed from one sector to its adjacent sector is given by
\begin{equation}
\overline{R} = \widetilde{T}^{n-1,n} R T^{n-1,n},
\label{Transformed_R_matrix}
\end{equation}
where $\widetilde{T}^{n-1,n}$ is the transpose of $T^{n-1,n}$.

Equations (\ref{R_matrix}) and (\ref{Transformed_R_matrix}) can be used to 
propagate the $R$ matrix sector by sector. Such a propagation starts at 
$r=a$, where the $R$ matrix of the field-free hydrogen atom can be calculated 
using Seaton's code~\cite{Seaton2002}, and stops in the asymptotic region with 
$r=b$. 

Mota-Furtado and O'Mahony~\cite{OMahony2007} actually provided another scheme 
to propagate the radial wave functions or the $R$ matrices. Their scheme adopts the 
global sector $R$ matrices derived in Ref.~\cite{Stechel} and denoted by 
${\mathcal R}_i^n$ in the present paper. The radial 
wave function and its derivative on the boundaries of the first sector and the 
$n$th sector are related by
\begin{equation}
F(a) = {\mathcal R}_2^nF^{\prime}(a_n) - {\mathcal R}_1^nF^{\prime}(a);
\label{Fmatrix_1_global}
\end{equation}
\begin{equation}
F(a_n) = {\mathcal R}_4^nF^{\prime}(a_n) - {\mathcal R}_3^nF^{\prime}(a).
\label{Fmatrix_2_global}
\end{equation}
We note that the sector-by-sector propagation is hidden in the global 
sector $R$ matrices. From the above two equations, one can obtain an expression 
to relate the $R$ matrices on the boundaries of the first sector and the final sector,
\begin{equation}
R(b)={\mathcal R}_4^n-{\mathcal R}_3^n[{\mathcal R}_1^n+
R(a)]^{-1}{\mathcal R}_2^n.
\label{R_matrix_global}
\end{equation}
In the current work, we used the scheme with the global sector $R$ matrices as given in 
Eq.~(\ref{R_matrix_global}) to propagate the $R$ matrix.

\subsection{Two-dimensional matching in the asymptotic region}
Since the electron motion in $\rho$ is bound, the Coulomb potential 
for large $z$ can be expanded into a series,  
\begin{equation}
-\frac1{\sqrt{z^2+\rho^2}} = - \frac1z + \frac{\rho^2}{2z^3} + 
O(\frac1{z^5})  + \cdot\cdot\cdot  
\label{Coulomb_potential_expanded}
\end{equation}
The Hamiltonian as specified by Eq.~(\ref{full_Hamiltonian}) becomes 
separable in cylindrical coordinates as $z\to\infty$. It is written as 
\begin{equation}
\widehat{H} = -\frac12\frac{d^2}{dz^2} - \frac1z + \frac{\rho^2}{2z^3}  + 
O(\frac1{z^5}) + \widehat{H}_L,
\label{Hamiltonian_cylindrical_coordinate}
\end{equation}
where $\widehat{H}_L$ is the Hamiltonian for the Landau states $\Phi_i(\theta,\phi)$. 
The potential (\ref{Coulomb_potential_expanded}) can be approximated by $-1/z$ 
for sufficiently large $z$, where the $j$th linearly independent solution of the 
Hamiltonian system (\ref{Hamiltonian_cylindrical_coordinate}) is written as 
\begin{equation}
{\Psi}_j=\sum_i \Phi_i(\rho,\phi) [s_i(z)\delta_{ij}+c_i(z)K_{ij}].
\label{linearly_independent_solution_in_cylindrical_coordinate}
\end{equation}
Here $s_i(z)$ and $c_i(z)$ are the energy-normalized regular and irregular Coulomb 
functions defined by Seaton~\cite{Seaton}, while $K_{ij}$ are the matrix elements 
of the reactance matrix~$K$, which is determined by a two-dimensional matching 
procedure. The functions $s_i(z)$ and $c_i(z)$ are calculated using the code written in Ref.~\cite{Zhao2016}. 
We performed the matching on an arc at $r=b$. 

Note that an extra index $j$ was added in $F_{\lambda}$ and $\Psi$ to label the solution specified 
in Eq.~(\ref{linearly_independent_solution_in_spherical_coordinate}). We set 
Eq.~(\ref{linearly_independent_solution_in_spherical_coordinate}) with $j$ added and 
Eq.~(\ref{linearly_independent_solution_in_cylindrical_coordinate}) equal to each other 
and then project both sides of the resulting equation onto 
$\varphi_{\lambda}(r; \theta,\phi)$.  This yields 
\begin{equation}
\frac1rF_{\lambda j}(r)=\sum_i \left[ P_{\lambda i}(r)\delta_{ij}+Q_{\lambda i}(r)K_{ij}\right],
\label{projection_both_sides}
\end{equation}
with 
\begin{equation}
P_{\lambda i}(r) = r \int \varphi_{\lambda}^*(r; \theta,\phi)\Phi_i(\rho,\phi)s_i(z)d\Omega 
\label{P_matrix_element}
\end{equation}
and
\begin{equation}
Q_{\lambda i}(r) = r \int \varphi_{\lambda}^*(r; \theta,\phi)\Phi_i(\rho,\phi) c_i(z)d\Omega.
\label{Q_matrix_element}
\end{equation}
The derivatives of $P_{\lambda i}(r)$, $Q_{\lambda i}(r)$, and $F_{\lambda j}(r)$ with 
respect to $r$ can be worked out from the above three equations. Using the definition 
of the $R$ matrix, one obtains  
\begin{equation}
R = [P+QK][P'+Q'K]^{-1}
\label{R_matrix_cylindrical_coordinate},
\end{equation}
where the $R$ matrix at $r = b$ is calculated from Eq.~(\ref{R_matrix_global}). 
Rearranging the above equation yields 
\begin{equation}
K = -[RQ'-Q]^{-1}[RP'-P].
\label{K_matrix}
\end{equation}
We note that our two-dimensional matching procedure given above is 
similar to that employed by Watanabe and Komine~\cite{Watanabe}, but it
differs from that of Mota-Furtado and O'Mahony~\cite{OMahony2007}. 

\subsection{Differential oscillator strength and cross sections}
The differential oscillator strength for the photo\-ionization process from an initial 
state $\Psi_i$ with energy $\varepsilon_i$ to a final state $\Psi_j^-$ with energy 
$\varepsilon$ is given by 
\begin{equation}
\frac{df_{j,i}}{d\varepsilon}=2(\varepsilon - \varepsilon_i) |\langle \Psi_j^-
|D|\Psi_i \rangle|^2,
\label{oscillator}
\end{equation}
where $D$ is the electric dipole operator in the length gauge and $\Psi_j^-$ is the energy-normalized 
incoming wave function. If the influence of the magnetic field on the atom is 
negligibly small in the region close to the nucleus, $\Psi_j^-$ in this region reduces to
\begin{equation}
\Psi_j^- = \sum_{\ell} \frac{1}{r} F^-_{\ell j}(r)Y_{\ell m}(\theta,\phi),
\label{incoming_wave_function_near_nuclus}
\end{equation}
where $F^-_{\ell j}$ can be written as a product of the field-free solution, which 
is the diagonal matrix with elements $S_{ij}=s_{i}\delta_{ij}$, and a constant 
denoted by $A^-$, which should be determined by means of the asymptotic 
solution. From Eq.~(\ref{Fmatrix_1_global}), we have
 \begin{equation}
[S(a)+{\mathcal R}_1^NS'(a) ] A^- = {\mathcal R}_2^N F^{\prime -}(b),
\label{Amatrix}
\end{equation}
where the $F^{\prime -}(b)$ are the energy-normalized incoming asymptotic solutions, which are
constructed from the standing-wave solutions given in Sec.~\ref{sec:Theory}~C. The physical 
reactance matrix is recovered using Seaton's quantum-defect theory~\cite{Seaton} 
from the reactance matrix~$K$. The coefficients $A^-$ are obtained by solving 
the above equation. Note that they are both field- and energy-dependent. 

Using the obtained coefficients $A^-$, Eq.~(\ref{oscillator}) is  
rewritten as 
\begin{equation}
\frac{df_{j,i}}{d\varepsilon}=|A^-_{j}|^2  \frac{f_{j,i}}{d\varepsilon}{\bigg|}_{B=0},
\label{oscillator_with_Amatrix}
\end{equation}
where $\frac{\displaystyle f_{j,i}}{\displaystyle d\varepsilon}{\big|}_{B=0}$ represents the field-free 
differential oscillator strength. The total differential oscillator strength is an observable 
quantity.  It is obtained by summing the differential oscillator strengths for photo\-ionization 
to the individual states according to
\begin{equation}
\frac{df_{i}}{d\varepsilon} = \sum_j \frac{f_{j,i}}{d\varepsilon}.
\label{total_oscillator}
\end{equation}
The photo\-ionization cross section is related to the total differential oscillator strength by
\begin{equation}
\sigma_i(\varepsilon) =2 \pi^2 \alpha \frac{df_{i}}{d\varepsilon},
\label{cross_section}
\end{equation}
where $\alpha$ is the fine-structure constant.
\section{Results and discussion}\label{sec:Results}
\begin{figure}
\includegraphics[width=0.95\columnwidth]{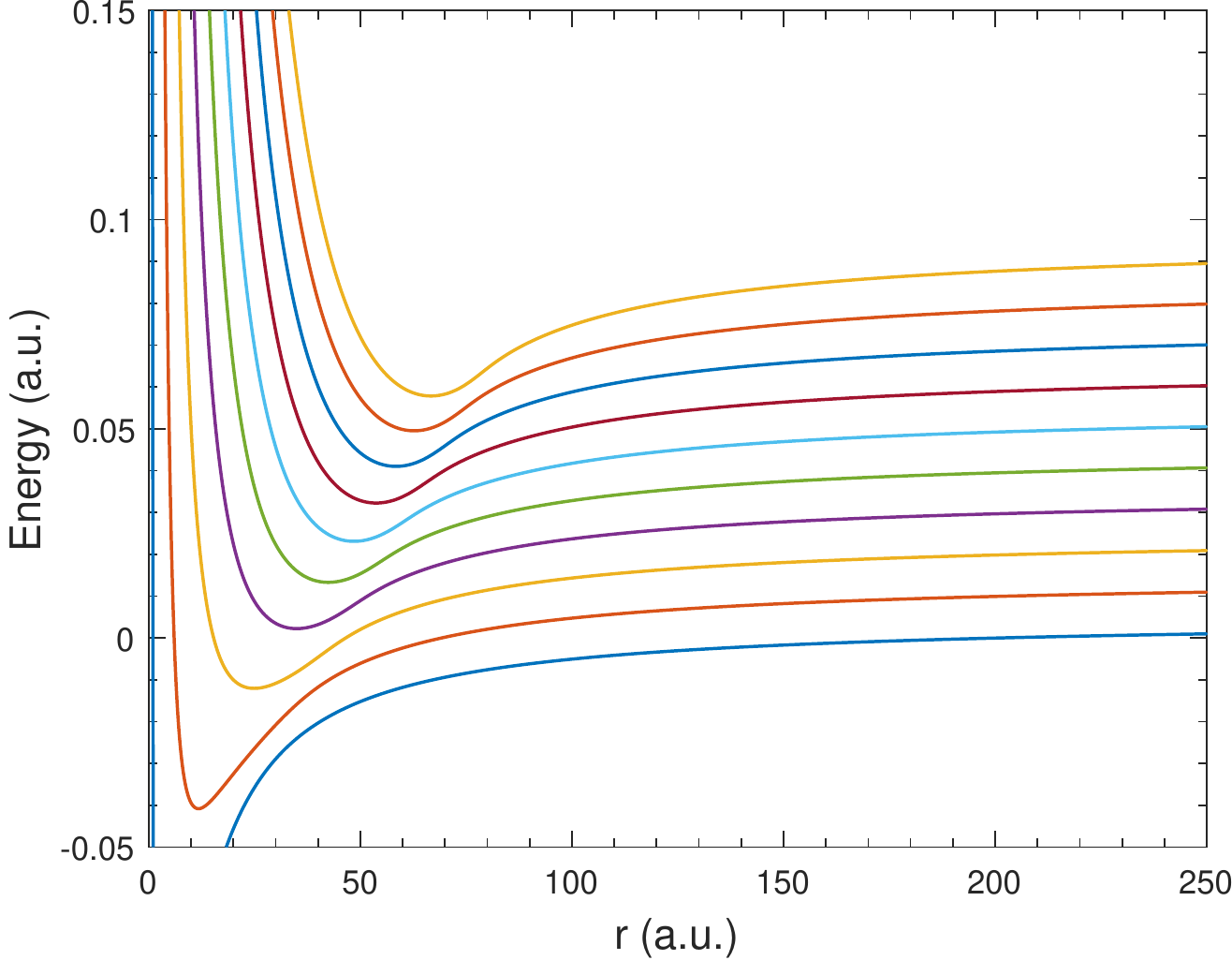}
\caption{Lowest ten adiabatic eigenvalue curves as a function of radius for the atomic state 
\hbox{$m^{\pi_z} = 0^{-}$} at $\gamma = 0.01$~a.u.} 
\label{adiabatic_curve}
\end{figure}

This section presents results of our investigation regarding photo\-ionization of hydrogen atoms in a 
white-dwarf-strength magnetic field, as obtained with the adiabatic-basis-expansion method outlined 
above. For each atomic state $m^{\pi_z}$ with a given magnetic field, we diagonalize 
the adiabatic Hamiltonian ${\widehat H}_{ad}$ in a basis set of spherical harmonics as 
a function of the radius and then plot the resulting adiabatic eigenvalue curves. We utilize these 
curves to determine the asymptotic radius~$b$, where a two-dimensional matching to the 
asymptotic solutions is performed. Figure~\ref{adiabatic_curve} depicts the lowest ten adiabatic 
eigenvalue curves for the atomic state $m^{\pi_z} = 0^{-}$ at \hbox{$\gamma = 0.01$ a.u.} 
These curves begin to display an equal energy spacing of Landau levels from $r \approx 200$~a.u.\ onwards. 
We vary $r$ around 200~a.u.\ to check the convergence of the predicted photo\-ionization oscillator strengths 
and finally fix the outer radius~$b$. 

\begin{figure}
\includegraphics[width=0.95\columnwidth]{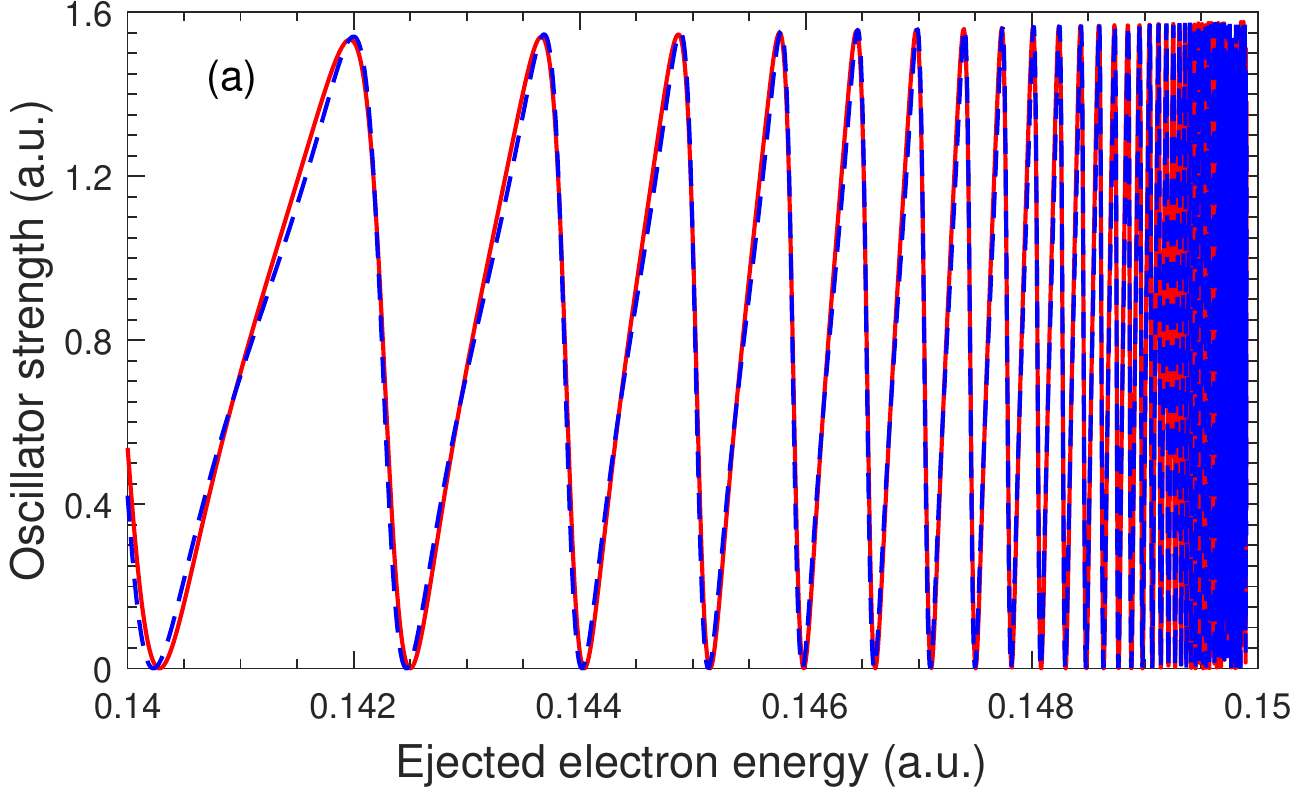}
\includegraphics[width=0.95\columnwidth]{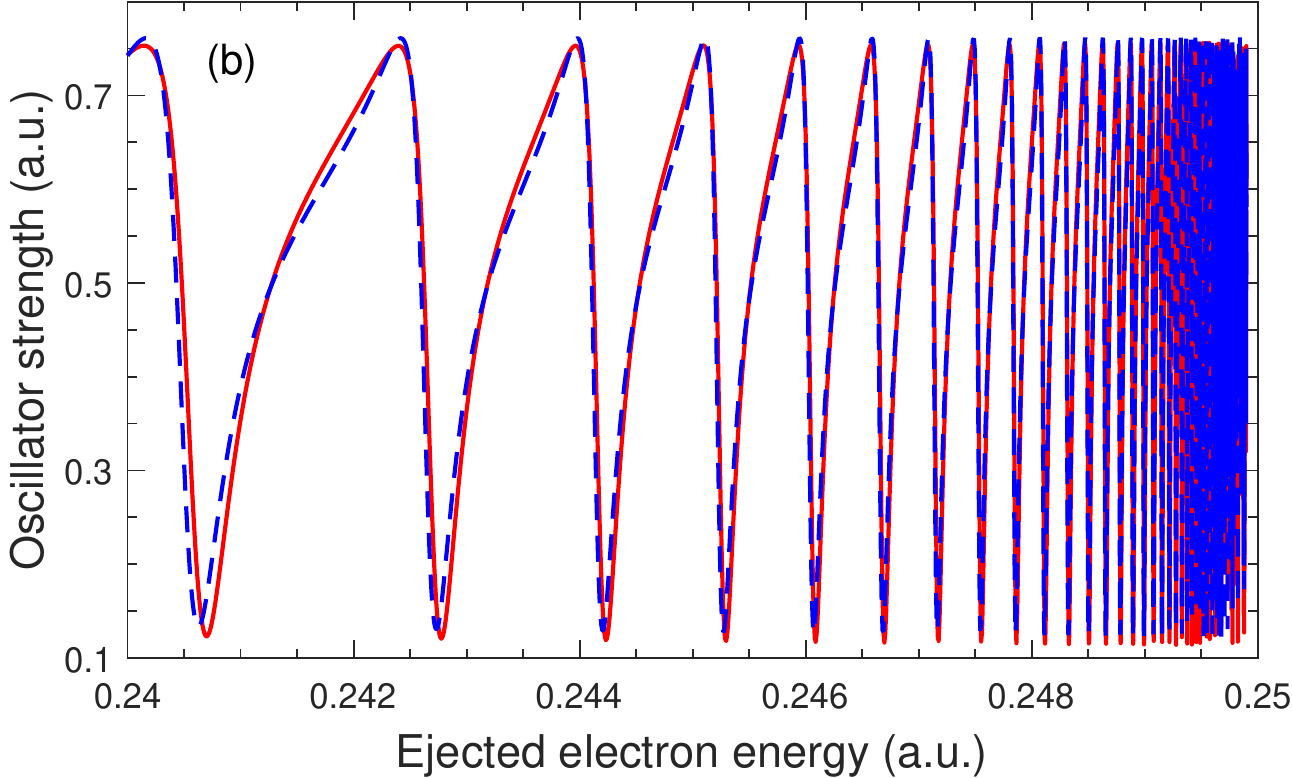}
\caption{Comparison of the Rydberg spectrum for photo\-ionization into the final 
continuum state $m^{\pi_z} = 0^{-}$ from the ground state of hydrogen atoms 
in a magnetic field with \hbox{$\gamma = 0.1$~a.u.}, just below the second~(a) and third~(b) 
Landau thresholds. The red solid and blue dashed curves represent the results 
from the current adiabatic-basis-expansion method and the coupled-channel theory~\cite{Zhao2016}, respectively. } 
\label{spectrum0.1au}
\end{figure}

A spectrum for photo\-ionization into the final continuum state $m^{\pi_z} = 0^{-}$ from 
the ground state at \hbox{$\gamma = 0.1$ a.u.}\ was presented by Mota-Furtado and 
O'Mahony~\cite{OMahony2007}. We recalculated this spectrum with energies covering the 
range from the first up to the third Landau thresholds using the current adiabatic-basis-expansion 
method, while our asymptotic solutions differ from those of Ref.~\cite{OMahony2007}. 
Nevertheless, one would expect that our calculations produce similar spectra to 
theirs. Unfortunately, we do not have access to their actual data, and therefore no detailed 
comparison can be made. However, visual inspection of their Fig.~6 indeed suggests very good agreement.
Furthermore, we performed a comparison with the available spectral 
data from the coupled-channel theory~\cite{Zhao2016} and found excellent agreement 
over the entire energy range covered. Figure~\ref{spectrum0.1au} displays the part of the 
Rydberg spectrum right below the second and third Landau thresholds from the two 
methods to illustrate this excellent agreement. 

This agreement shows the reliability of the current theoretical 
method and provides confidence in revisiting the problem of photo\-ionization of 
atomic hydrogen in a magnetic field of 2,000~T with the adiabatic-basis-expansion 
method.  Note that the continuum spectra at 2,000~T reported in the literature are 
in significant disagreement with each other~\cite{Alijah,Wang,Zhao2016}. 

We first calculated oscillator strengths for photo\-ionization into the final continuum state 
$m^{\pi_z} = 1^+$ from the ground state of atomic hydrogen in a magnetic field 
of 2,000~T. We sought the outer radius $b$ by plotting the adiabatic eigenvalue 
curves for the atomic state $m^{\pi_z} = 1^+$ and found $b\approx 220$~a.u.\ 
from the asymptotic behavior of these curves. As mentioned above, the convergence 
of the predicted oscillator strengths was checked by varying $b$ in the vicinity of 220~a.u. 
The stability of the results was checked further by varying other parameters, such as the inner 
radius~$a$, the number of channels involved, and the number and size of the sectors. 

\begin{figure}
\includegraphics[width=0.95\columnwidth]{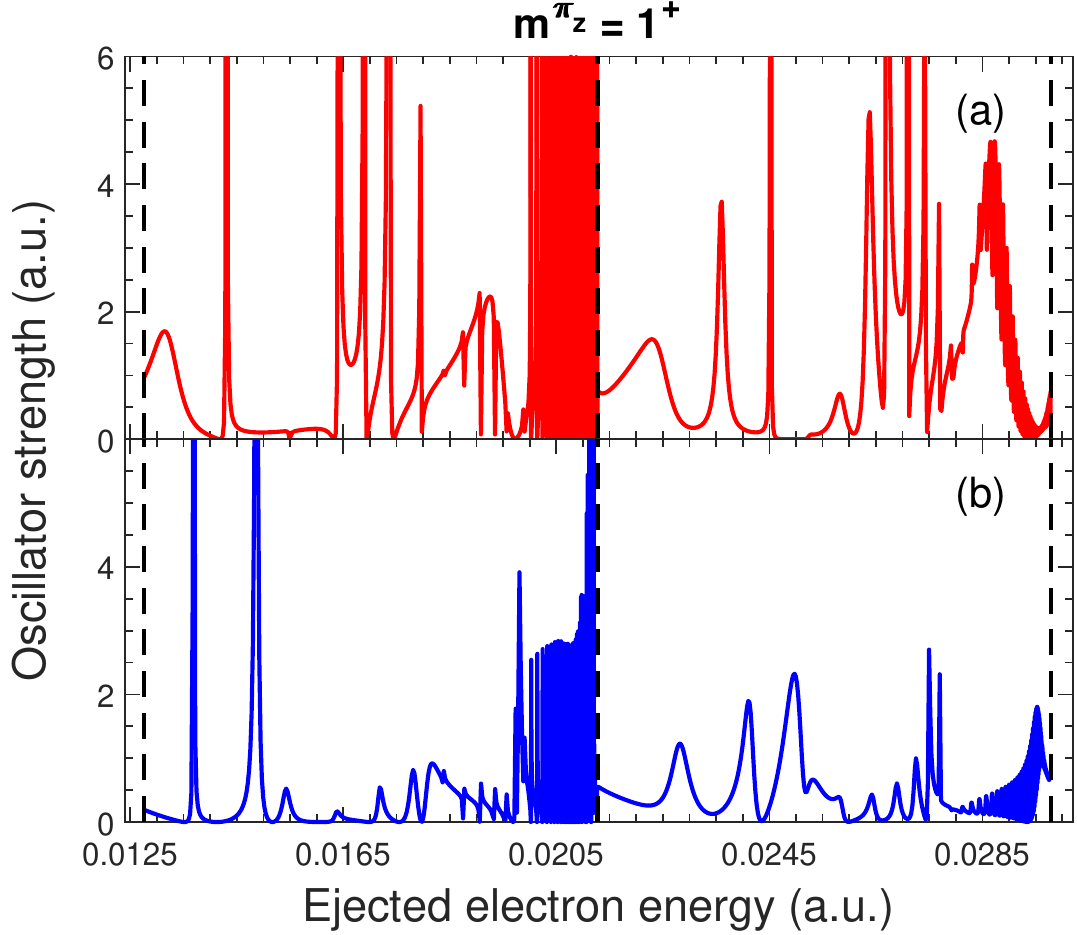}
\includegraphics[width=0.95\columnwidth]{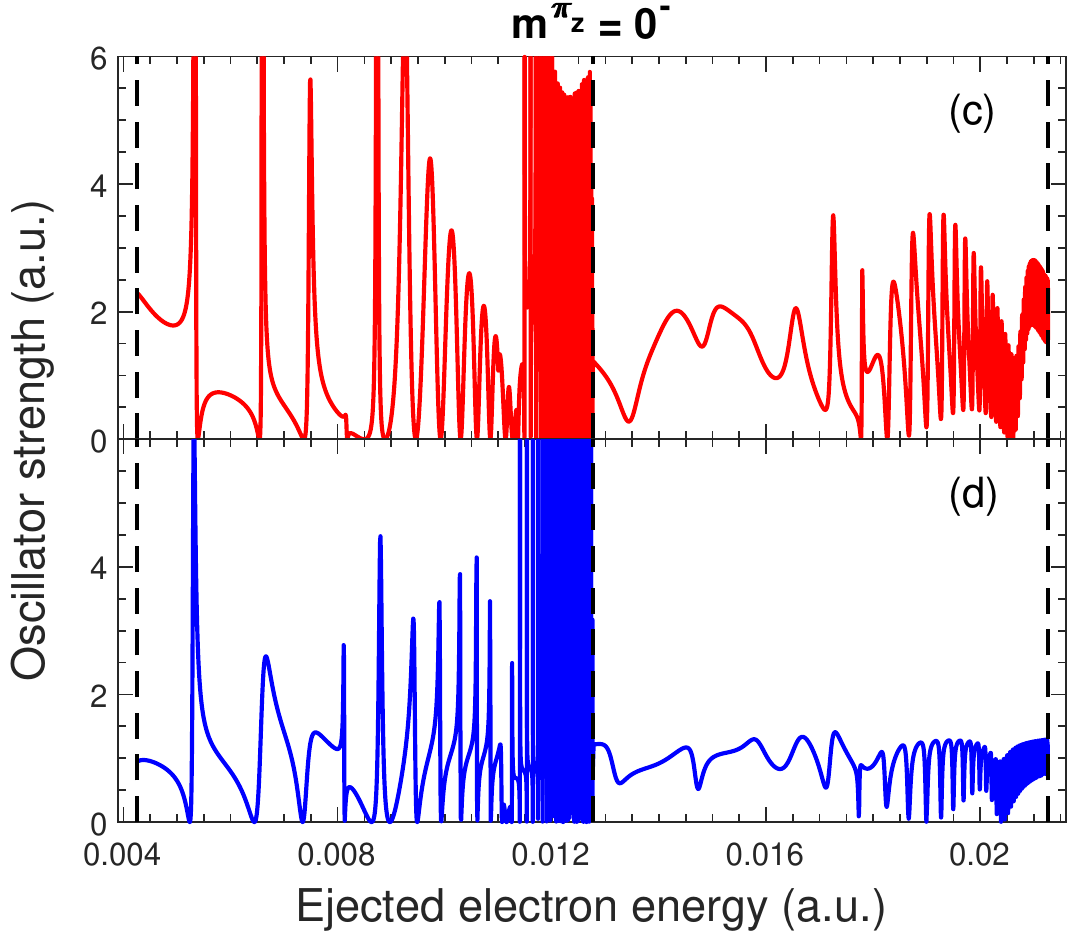}
\caption{Comparison of photo\-ionization spectra for ground-state hydrogen atoms in a 
magnetic field of \hbox{$B$ = 2,000~T} obtained with the current adiabatic-basis-expansion 
method (a,c) and the coupled-channel theory (b,d). The two photo\-ionization 
processes into the various final continuum states are labeled $m^{\pi_z}=1^+$ 
and $m^{\pi_z}=0^-$, respectively. The ejected-electron energies cover the range 
from the first to the third Landau thresholds, indicated by the dashed vertical lines,  
for both photo\-ionization processes.} 
\label{spectrum2000T}
\end{figure}

The calculated spectrum from the first to the third Landau ionization thresholds is shown 
in Fig.~\ref{spectrum2000T}~(a). The spectrum in the same energy region was also 
calculated using the coupled-channel theory~\cite{Zhao2016} and is plotted in 
Fig.~\ref{spectrum2000T}~(b) for comparison. The results from these two methods 
are in qualitative agreement, but pronounced discrepancies between the two spectra 
are clearly visible in the details. A broad resonance structure underlying the Rydberg 
resonance peaks right below the third Landau threshold can be seen in both of these 
spectra, but their positions, widths, and heights differ. The broad resonance was 
attributed by Alijah {\it et al.}~\cite{Alijah} to the downward-shifted $n = 8$ state of 
the fourth Landau channel. 

The level of disagreement displayed in Figs.~\ref{spectrum2000T}~(a,b) was unexpected. 
On the other hand, the current calculations do not 
reproduce the part of the spectrum right below the third Landau threshold reported 
by Alijah {\it et al.}~\cite{Alijah} either. Finally, our Rydberg spectrum below the 
third Landau threshold also differs from that of Wang and Greene~\cite{Wang}. 
Using the \hbox{$R$-matrix} approach, which is constructed within the framework 
of multi\-channel quantum-defect theory, to analyze the close-coupling calculations 
of Alijah {\it et al.}~\cite{ Alijah} in detail, Wang and Greene concluded that the 
box size $z_0 = 50$~a.u used by Alijah {\it et al.} is too small.  

We then calculated photo\-ionization into the final continuum state $m^{\pi_z} = 0^-$ 
from the ground state of hydrogen atoms in a magnetic field of 2,000~T with both 
the current method and the coupled-channel theory~\cite{Zhao2016}.  The obtained 
oscillator strengths are displayed as a function of ejected electron energy in 
Figs.~\ref{spectrum2000T}~(c) and~(d). We again expected better agreement in the details 
than what is observed in these figures.  Only qualitative agreement between 
the results from the two theoretical methods is still visible over the entire energy region 
covered.

\begin{figure}
\includegraphics[width=0.95\columnwidth]{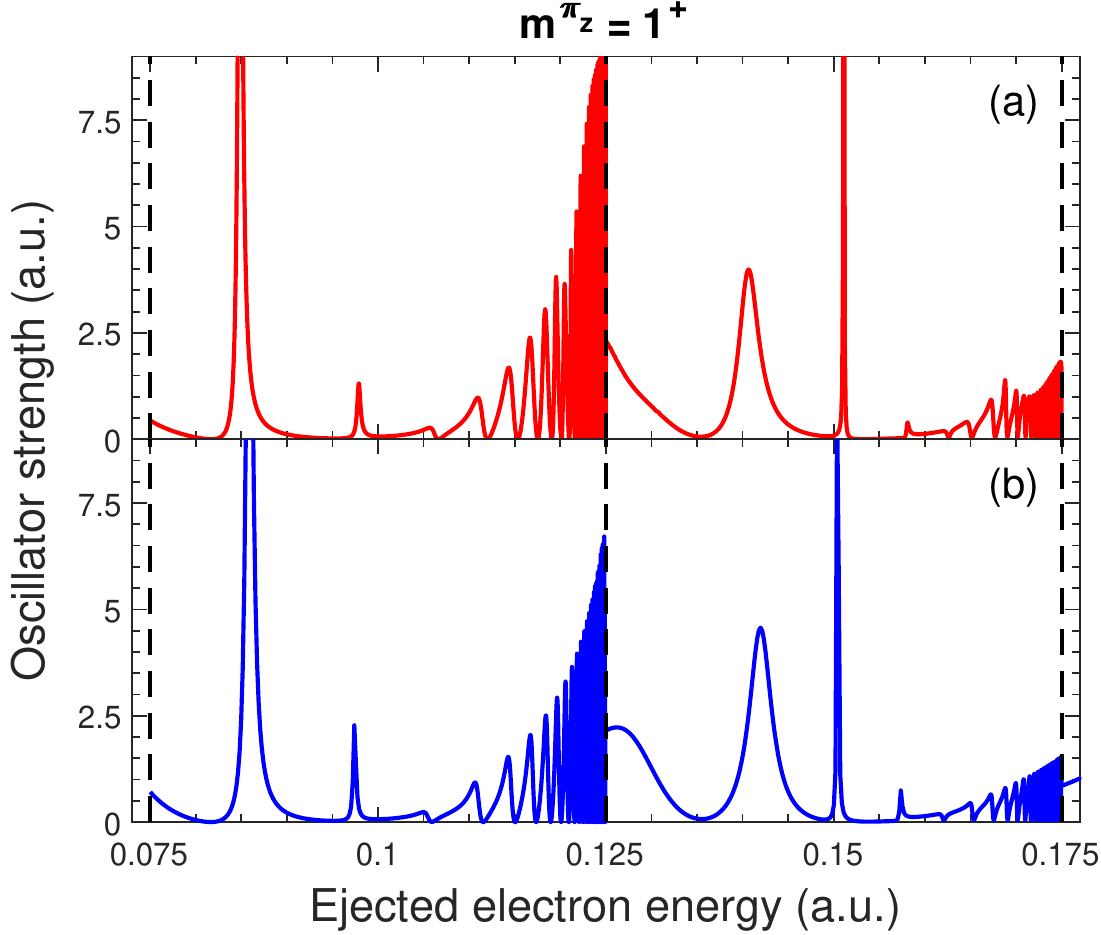}
\includegraphics[width=0.95\columnwidth]{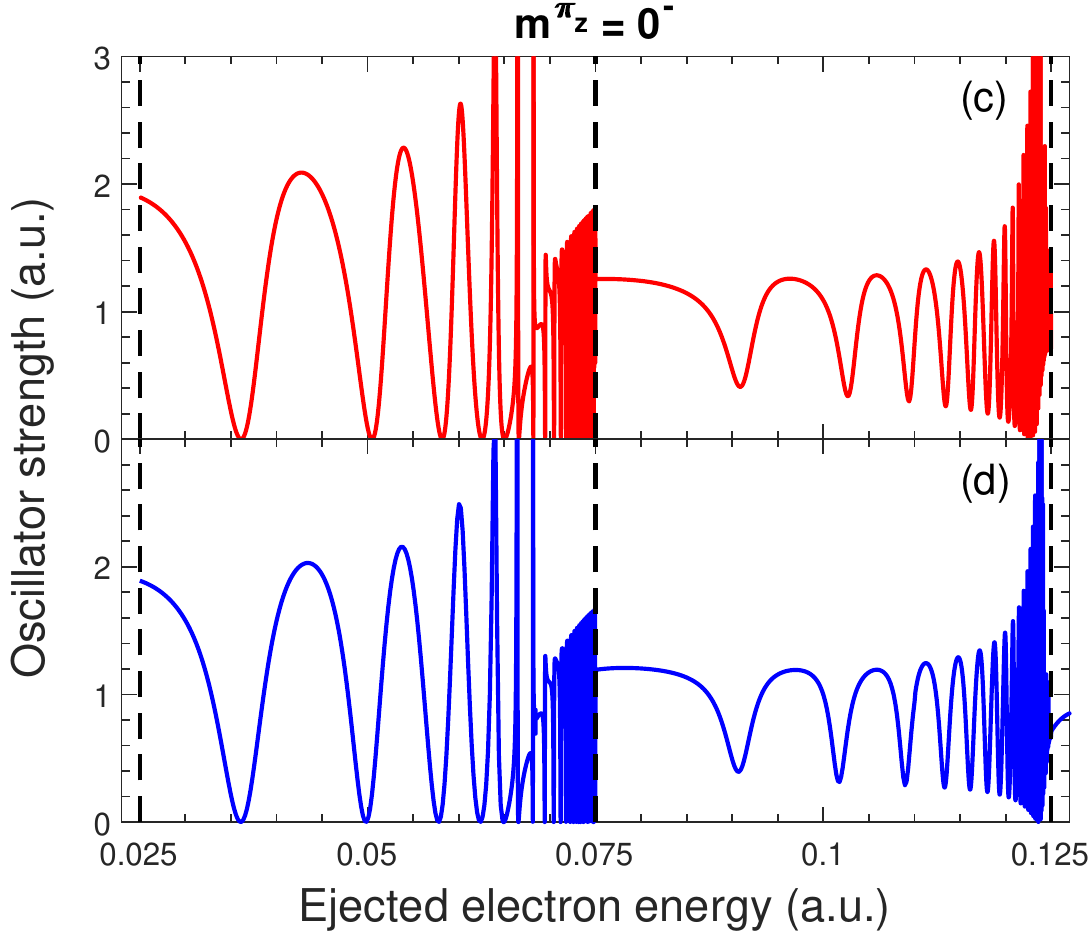}
\caption{Same as Fig.~\ref{spectrum2000T}, but with $\gamma = 0.05$~a.u.}
\label{spectrum0.05au}
\end{figure}

In order to better understand the discrepancies of the Rydberg spectra shown in Fig.~\ref{spectrum2000T}, 
we calculated photo\-ionization spectra of hydrogen atoms in 
magnetic fields with different field strengths. Figure~\ref{spectrum0.05au} displays 
our calculated oscillator strengths in a magnetic 
field with \hbox{$\gamma$ = 0.05~a.u.}\ as a function of the ejected-electron energy. We assume 
that magnetized hydrogen atoms in the ground state are irradiated by beams of 
circularly and linearly polarized light, respectively, and then ionized into the two final 
continuum states $m^{\pi_z} = 1^+$ and $0^-$.  The results from 
the coupled-channel theory are also shown in this figure for comparison.  Good 
agreement between the results from the two methods is evident for both photo\-ionization 
processes, although some small discrepancies exist. 
One readily sees, for example, small shifts of the resonance positions near 
$\varepsilon = 0.078$, 0.14, and 0.15~a.u., as well as a minor discrepancy in the resonance width near 
$\varepsilon$ = 0.078~a.u.

\begin{figure}
\includegraphics[width=0.95\columnwidth]{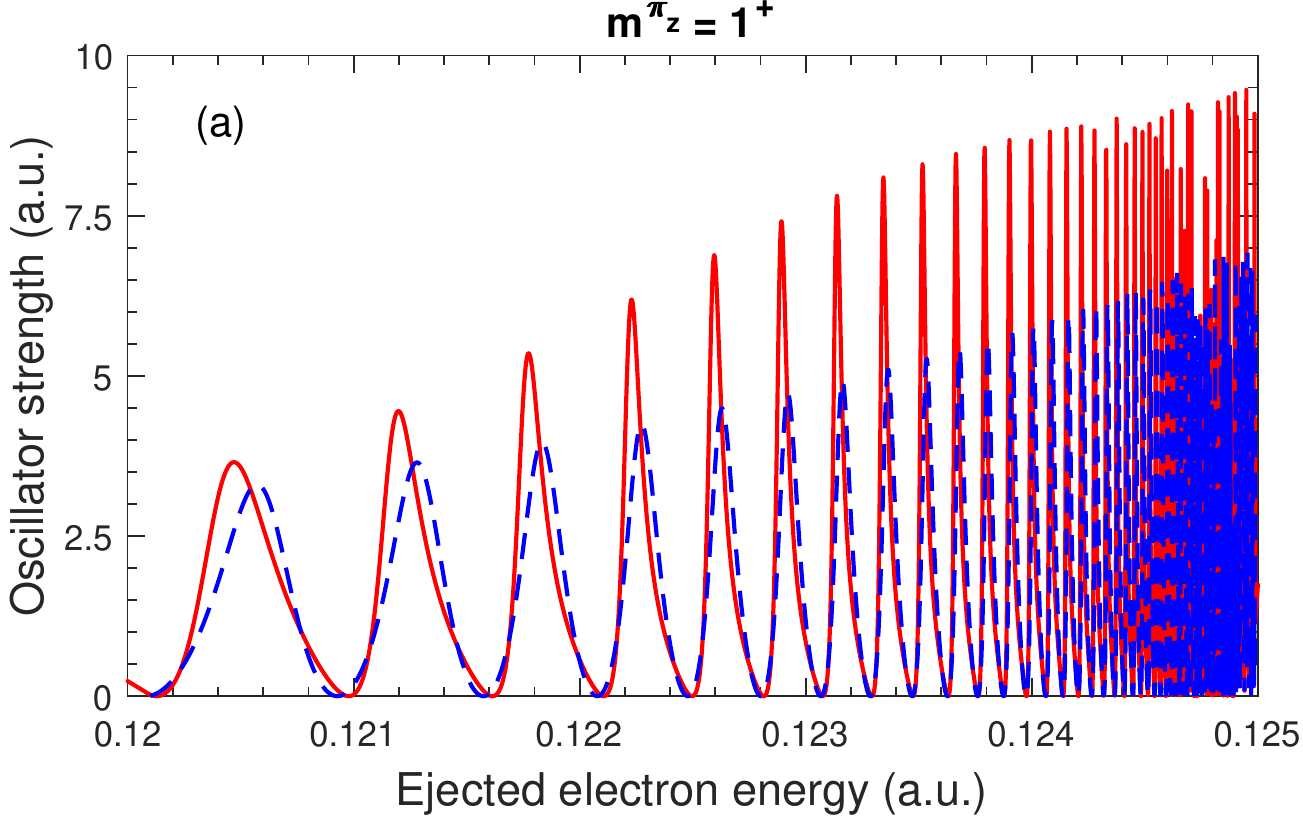}
\includegraphics[width=0.95\columnwidth]{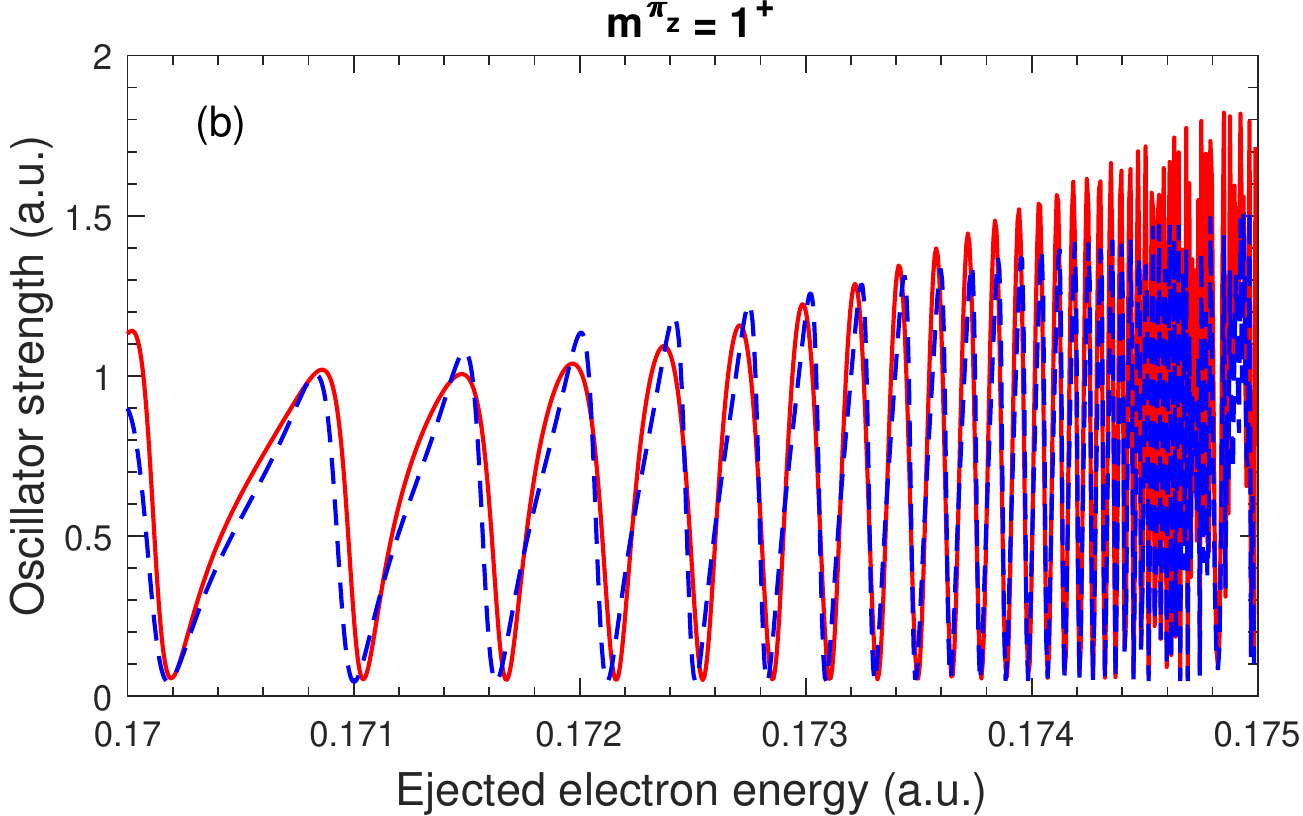}
\includegraphics[width=0.95\columnwidth]{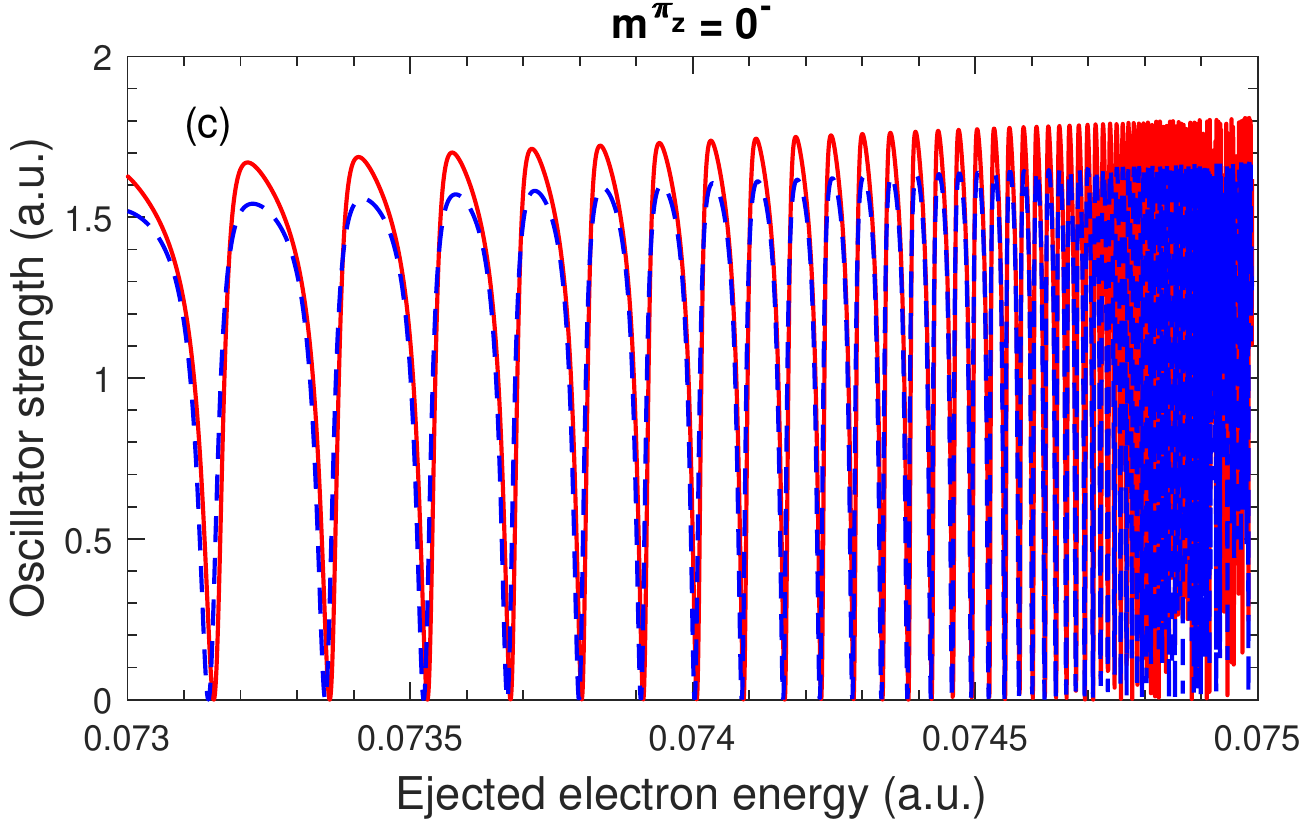}
\includegraphics[width=0.95\columnwidth]{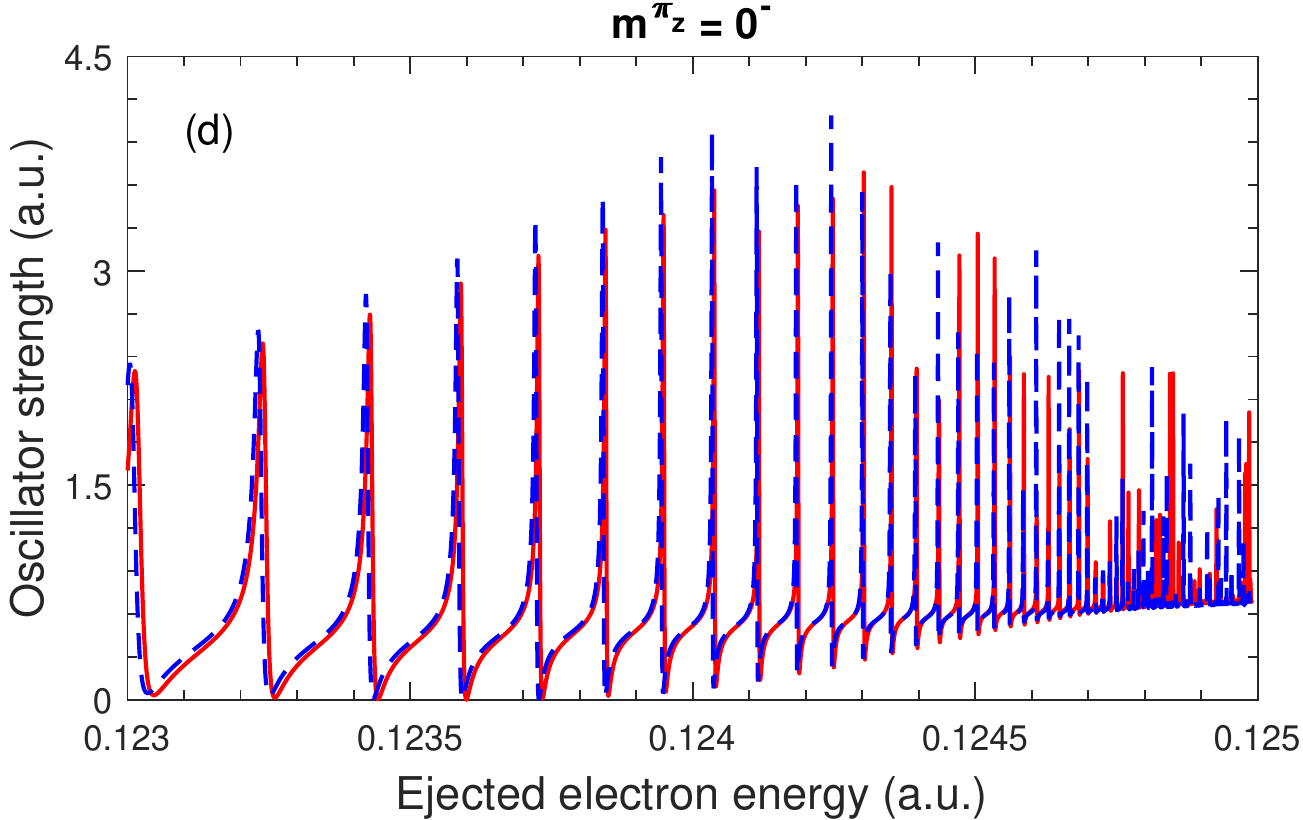} 
\caption{Detailed comparison of Rydberg spectra, as given in Fig.~\ref{spectrum0.05au}, 
between the current adiabatic-basis-expansion method (red solid curves) and the 
coupled-channel theory (blue dashed curves). The relevant energy regions are just 
below the second~(a,c) and third~(b,d) Landau thresholds.}
\label{spectrum0.05au_Rydberg}
\end{figure}

The parts of the spectrum right below the Landau thresholds displayed in Fig.~\ref{spectrum0.05au} 
are not sufficiently resolved. It is, however, helpful to display a detailed comparison of these
parts in order to understand the discrepancies between the Rydberg spectra in 
Fig.~\ref{spectrum2000T}.  Such a detailed comparison right below the second and 
third Landau thresholds is presented in Fig.~\ref{spectrum0.05au_Rydberg}. The resonances 
associated with high-lying Rydberg states are becoming increasingly narrow 
as the ejected electron energies approach the Landau threshold. Discrepancies 
in the heights of these Rydberg resonances obtained by the two theoretical methods can 
be seen for the two photo\-ionization processes into the two final continuum states 
$m^{\pi_z} = 1^+$ and $0^-$, but their positions, widths, and overall energy dependence are in good 
agreement with each other. 

\begin{figure}
\includegraphics[width=0.85\columnwidth]{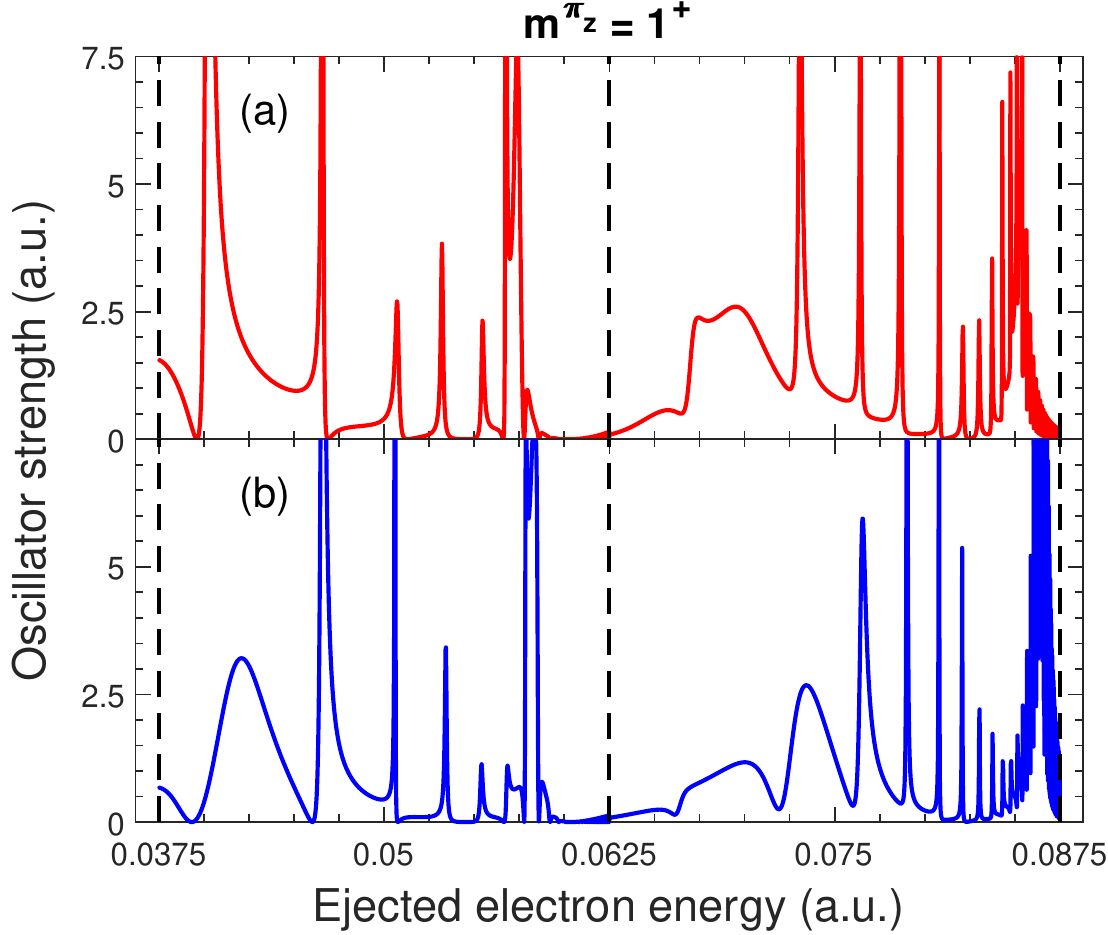}
\includegraphics[width=0.85\columnwidth]{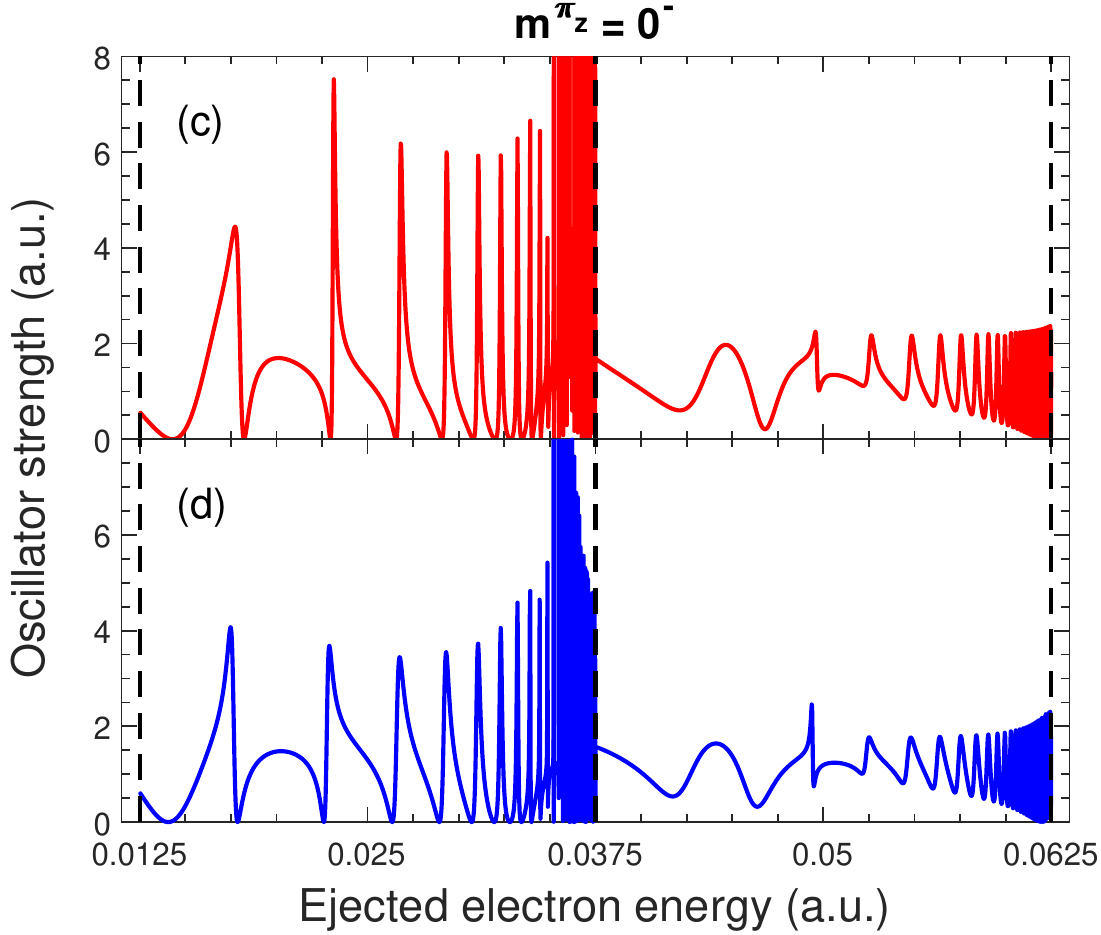}
\caption{Same as Fig.~\ref{spectrum0.05au}, but with $\gamma = 0.025$~a.u.}
\label{spectrum0.025au}
\end{figure}

Finally, the spectra for photoionization into the two final continuum states $m^{\pi_z} = 1^+$ and
$0^-$ from the ground state at a magnetic field with $\gamma$ = 0.025~a.u.\ were again calculated
using both the current adiabatic-basis-expansion method and the coupled-channel theory.
The obtained oscillator strengths as a function of ejected-electron energies covering
the region from the first to third Landau threshold are presented in Fig.~\ref{spectrum0.025au}.
Good agreement between these two approaches can be seen for photoionization into
$0^-$ over the entire energy region. In particular, we performed a detailed comparison of the
parts of the spectra right below the second and third Landau thresholds, as done in
Fig.~\ref{spectrum0.05au_Rydberg}.  We see that the resonance positions, widths, and
overall energy dependence from the two methods are in good agreement. For photoionization
into the final state $1^+$, the spectra from the two methods show good agreement in the overall
energy dependence, but there are visible shifts in the resonance positions. We note that
Wang and Greene~\cite{Wang} also presented their spectra for this magnetic field strength. Our
results are found to be in qualitative agreement with theirs, but there are discrepancies in the details.

Note that Wang and Greene~\cite{Wang} expressed caution regarding 
the reliability of their results at relatively low field strengths. Given the fact that the 
coupled-channel theory adopts the cylindrical coordinate system, it should  
be most appropriate for relatively high field strengths. The current results illustrate that this method 
may approach its limit of validity for magnetic fields around \hbox{$\gamma = 0.01$~a.u.} On the contrary, 
the adiabatic-basis-expansion method was shown to be reliable for low field 
strengths when comparing its predictions with experimental results at 6.1143~T~\cite{OMahony}. 
We therefore believe that the spectra for photo\-ionization of magnetized 
hydrogen atoms at 2,000~T, as calculated in the present work with the adiabatic-basis-expansion method,
are more accurate than those from both the coupled-channel theory~\cite{Zhao2016} and the approach
used by Wang and Greene~\cite{Wang}.     

\section{Summary and conclusions}\label{sec:Summary}
Since there exist pronounced discrepancies in the predicted positive-energy 
spectra of atomic hydrogen in a white-dwarf-strength magnetic field obtained by a variety 
of theoretical approaches reported in the literature, we revisited the problem using the 
adiabatic-basis-expansion method developed by Mota-Furtado and O'Mahony~\cite{OMahony2007}, 
with one significant modification. Specifically, we adopted a different two-dimensional 
matching procedure to exact the reactance matrix in order to simplify the related calculations.  
Our test calculations show that such a procedure does not cause any significant numerical inaccuracy. 

We then performed a comparative study between the current adiabatic-basis-expansion method 
and our previously developed coupled-channel theory~\cite{Zhao2016}. Our calculated 
positive-energy spectra were also compared to those from other theoretical approaches. 
A detailed analysis suggests that the adiabatic-basis-expansion method can produce 
more accurate positive-energy spectra than all the other reported approaches for 
relatively low field strengths. While we hope that the current study finalizes the problem 
of photo\-ionization of atomic hydrogen in a white-dwarf-strength magnetic field, we would 
welcome additional studies from other groups. 

\section*{Acknowledgements}
This work was supported by the National Science Foundation of China under Grant No.~11974087 
(L.B.Z.), the Program for Science and Technology Innovation Talents in Universities of the Henan 
Province of China under grant No.~19HASTIT018 (K.D.W.), and the United States
National Science Foundation under Grant No.~PHY-1803844 (K.B.). 



\end{document}